\documentclass[aps, a4paper,superscriptaddress,floatfix, nofootinbib, showpacs, twocolumn, 10pt]{revtex4}
\usepackage{amsmath,amssymb,bm,natbib}
\usepackage{epsfig}
\usepackage{graphicx}
\usepackage{slashed}
\usepackage{color}
\newcommand{\beq}{\begin{equation}}
\newcommand{\eeq}{\end{equation}}
\newcommand{\bea}{\begin{eqnarray}}
\newcommand{\eea}{\end{eqnarray}}
\newcommand{\ba}{\begin{align}}
\newcommand{\ea}{\end{align}}
\newcommand{\bfig}{\begin{figure}}
\newcommand{\efig}{\end{figure}}

\newcommand{\as}{\alpha_s}
\newcommand{\wh}{\widehat}
\newcommand{\nn}{\nonumber}

\newcommand{\eqn}[1]{(\ref{#1})}

\newcommand\fverb{\setbox\pippobox=\hbox\bgroup\verb}
\newcommand\fverbdo{\egroup\medskip\noindent%
                        \fbox{\unhbox\pippobox}\ }
\newcommand\fverbit{\egroup\item[\fbox{\unhbox\pippobox}]}
\newbox\pippobox

\begin{document}

\vskip1cm
\title{Determination of $\alpha_s(M_{\tau}^2)$ from 
Improved Fixed Order Perturbation Theory}
\author{Gauhar Abbas}
\affiliation{Centre for High Energy Physics,
Indian Institute of Science, Bangalore 560 012, India}
\author{B.Ananthanarayan}
\affiliation{Centre for High Energy Physics,
Indian Institute of Science, Bangalore 560 012, India}
\author{Irinel Caprini}
\affiliation{
Horia Hulubei National Institute for Physics and Nuclear Engineering,
P.O.B. MG-6, 077125 Magurele, Romania}

\begin{abstract}
We revisit the extraction of $\alpha_s(M_\tau^2)$ from the QCD
perturbative  corrections to the hadronic $\tau$ branching ratio, using an  improved 
fixed-order perturbation theory based on the explicit summation of all 
renormalization-group accessible logarithms, 
proposed some time ago in the literature. In this approach, the powers of the
 coupling in the expansion of the  QCD Adler function are multiplied by a set of 
functions $D_n$, which depend themselves on the coupling and can be written in a 
closed form by iteratively solving a sequence of differential equations. 
We find that the new expansion has an improved behavior in the complex energy plane 
compared to that of the standard fixed-order perturbation theory  (FOPT), 
and is similar but not identical to the contour-improved  perturbation theory 
(CIPT). With five terms in the perturbative expansion we obtain in the 
${\overline{\rm MS}}$ scheme $ \alpha_s(M_\tau^2)= 0.338 \pm 0.010$, 
using as input a precise value for the perturbative contribution to
the hadronic width of the $\tau$ lepton reported recently  in the literature.  
\end{abstract}
\maketitle

\section{Introduction} 

 The non-strange hadronic decays of the $\tau$ lepton provide one of the most 
precise determination of the strong coupling $\alpha_s$.
The recent calculation of the Adler function to four loops \cite{BCK08}, 
the same order to which the $\beta$-function of the renormalization-group (RG) 
equation is known \cite{LaRi,Czakon}, renewed interest in the 
determination of $\alpha_s(M_\tau^2)$ \cite{Davier2008}-\cite{DV}. 
The intriguing remark \cite{BeJa} that the inclusion of a  higher order term  
increased, instead of reducing,  the theoretical error on the resulting 
$\alpha_s(M_\tau^2)$   stimulated many investigations aimed at understanding this fact.

 The basic procedure involves the analytic continuation of the  
Adler function (the logarithmic derivative of the massless QCD polarization function) 
in the complex energy plane, where it can be calculated by the Operator Product
 Expansion (OPE).  The contribution of the higher dimensional 
terms (``power corrections") in the OPE to the $\tau$ hadronic width was evaluated and found to be 
quite small \cite{BrNaPi, Davier2006,Davier2008, BeJa, Pich2010, Pich_Muenchen}. Recently, the effect of the nonperturbative terms was investigated in a more general framework, which includes also  deviations of 
the true polarization function from the OPE description, especially near 
the timelike axis, {\em i.e.} violation  of quark-hadron duality
\cite{DV}. 

There are two competing versions of perturbation theory, 
the standard fixed-order perturbation theory (FOPT) and the RG-improved,
which in this context is also
known as contour-improved perturbation theory 
(CIPT)~\cite{Pivovarov:1991rh,dLP1}. 
Their predictions differ by about 0.02, which is at present the main part of the 
theoretical error on $\alpha_s(M_\tau^2)$ \cite{Bethke, BeJa, Pich_Manchester, 
Pich_Muenchen, Beneke_Muenchen}.  It should, however, be noted
that the issue of the separation of the perturbative and nonperturbative
parts is not completely settled, with a potential effect on the precision of the $\alpha_s$ predictions. For instance, analyses based on the moments of the spectral functions, either standard \cite{MaYa} or including possible duality violating contributions   \cite{DV}, suggest a different value for the nonperturbative contribution 
to the hadronic width 
compared to that obtained from previous studies \cite{dLP2, Davier2006, Davier2008, BeJa, Pich_Muenchen, Beneke_Muenchen}.

The investigation of the perturbative series of QCD in the context of 
the uncertainty in the extraction of $\alpha_s$ is of such great importance 
that its theoretical
aspects have been studied by several authors and various alternative
approaches have been proposed. They  include in general additional information 
about the series beyond the truncation order, known either from specific 
classes of Feynman diagrams or from RG invariance.   
Thus, a reordering of the standard contour-improved approach exploiting 
RG invariance was proposed in  \cite{CLMV1}, and a 
detailed analysis of the errors of various expansions
 has been performed in \cite{Menke}. 

A more radical modification was investigated in 
\cite{CaFi2009,CaFi_Manchester,CaFi2011}, where the 
available knowledge on the large-order behavior of 
the perturbative coefficients was exploited with mathematical 
techniques of accelerating the series convergence by means of 
conformal mappings \cite{CiFi,CaFi1998,CaFi2000,CaFi2001}. This led 
to a modified expansion in terms of a new set of functions, which
 have the advantage of sharing the known singularities of the expanded 
correlator in the coupling and the Borel complex planes. As argued in 
\cite{CaFi2011}, this expansion is particularly suitable in the 
contour-improved version, since it make a summation of both the 
running coupling and of the Feynman coefficients of the Adler function.
 Detailed numerical studies \cite{CaFi2009,CaFi2011} proved the good 
convergence properties of the latter expansion for a large class of 
physical models which simulate the known properties of the Adler function. 

In the light of the above, any fresh attempt to improve the understanding of 
 the properties of the perturbative expansion in the complex energy plane and the origin of the discrepancy in the coupling predictions would be welcome. With this motivation,  we consider in the present paper  a RG-improved expansion proposed in  \cite{Ahmady1,Ahmady2}, using a procedure originally advocated  in \cite{MaMi1,Maxwell, MaMi2}. The method is a generalization of the leading logarithms summation, in which terms in powers of the coupling constant 
and logarithms are regrouped, so that for a given order, the new expansion includes every 
term in the  perturbative series that can be calculated using the RG invariance. The method was applied to several correlators and observables, for instance the inclusive  decays of the $b$-quark \cite{Ahmady1} and the
hadronic cross section in $e^+e^-$ collisions \cite{Ahmady2}, where its main merit was proved to be a substantial reduction in sensitivity  to the renormalization scale. In the present paper we investigate the new expansion for the QCD Adler function in the complex energy plane and the determination of $\alpha_s$ from $\tau$ hadronic decays.  To our knowledge,  this problem was not investigated in full generality up to now.\footnote{The RG-summation discussed in \cite{MaMi1,Maxwell}
has been applied to the extraction to $\alpha_s$ from $\tau$ decays in \cite{MaMi2}, but only using the perturbation series to NNLO treated with Borel summation methods.}
 We shall refer to this scheme as ``improved FOPT" where the improvement
is implied only in the sense of capturing the RG-summation of the accessible
logarithms.  {\it A priori} is does not imply any other kind of improvement.

The plan of this paper is as follows: for completeness we briefly review in 
Sec. \ref{sec:Adler} the perturbative expansion of the Adler function and its 
connection to the hadronic decay width of $\tau$. In Sec. \ref{sec:RGR}, 
following Ref. \cite{Ahmady2}, we review the derivation of the new 
RG-improved expansion of the Adler function and give the  
corresponding expansion functions calculated to four loops.  For further applications of the method  it is useful to know  also the  higher expansion 
functions, which we have calculated in an analytic closed form by iteratively solving the 
relevant differential equations. As the general expressions are rather lengthy, 
we give in the Appendix simpler forms of the expansion functions up to $n=10$ obtained by inserting the numerical values of the known perturbative coefficients  of both the Adler and $\beta$-functions to four loops. The expressions are written in terms of the coefficients beyond four loops, which are not yet available from explicit calculations and are left arbitrary.  In Sec. \ref{sec:disc} we investigate the properties of the new expansion in the complex energy plane and compare it with the standard FOPT and CIPT, using in particular a physical model for the Adler function proposed in \cite{BeJa}.   In Sec. \ref{sec:alphas} we apply the FO expansion improved by RG-summation discussed in this paper to  a determination of $ \alpha_s(M_\tau^2)$,  using the phenomenological value of the perturbative QCD contribution to the  hadronic width of  $\tau$ estimated recently in \cite{Beneke_Muenchen, Pich_Muenchen}. Section \ref{sec:conc} summarizes our results and presents some conclusions.

\section{Adler function in perturbative QCD}\label{sec:Adler}

The Adler function plays a crucial role in the determination of  $\alpha_s(M_\tau^2)$ from hadronic $\tau$ decays.  The method is discussed in the seminal paper \cite{BrNaPi} and is reviewed in several recent articles \cite{Davier2008, BeJa, Pich_Manchester, Pich_Muenchen}. For completeness we give below a few details. 

The inclusive character of the total $\tau$ hadronic width makes possible  an accurate
calculation of the ratio  
\begin{eqnarray*}
R_\tau \,\equiv\,\frac{\Gamma[\tau^- \to \nu_\tau {\rm hadrons} \,  ]}{
\Gamma[\tau^- \to \nu_\tau e^- \bar \nu_e ]}.
\end{eqnarray*}  
Of interest is the Cabbibo allowed component which proceeds either through a 
vector or an axial vector current, since in this case the power corrections are
particularly suppressed.  On the theoretical side,  $R_\tau$ can be expressed in the 
form
\begin{equation}
\label{RtauVA}
R_{\tau} \,=\, \frac{N_c}{2}\,S_{\rm EW}\,|V_{ud}|^2\,\biggl[\,
1 + \delta^{(0)} + \delta_{\rm EW}' + \sum\limits_{D\geq 2} 
\delta_{ud}^{(D)} \,\biggr] \,,
\end{equation}
where $N_c =3$ is the number of quark colors,
$S_{\rm EW}$  and
$\delta_{\rm EW}'$  are electroweak corrections,
$\delta^{(0)}$ is the dominant perturbative QCD correction, and the $\delta_{ud}^{(D)}$ denote quark mass
and higher $D$-dimensional operator corrections (condensate contributions) arising in
the OPE. 

 Unitarity implies that the inclusive hadronic decay rate can be written as a weighted integral along the timelike axis of the spectral function of the polarization function  $\Pi^{(1+0)}(s)$,  where the superscript denotes the angular momentum. As shown in \cite{BrNaPi}, the analytic properties of the polarization function and the Cauchy theorem allow one to write  equivalently this quantity  as an integral along a contour in the complex
$s$-plane (chosen for convenience to be the circle $|s|=M_\tau^2$).  After an integration by parts, in our notation  the quantity of interest $\delta^{(0)}$ is expressed as the following contour integral: 
\begin{equation}
\label{del0}
\delta^{(0)}=\frac{1}{2 \pi i}\!\! \oint\limits_{|s|=M_\tau^2}\!\! \frac{d s}{s}
\left(1- \frac{s}{M_\tau^2}\right)^3 \left(1+\frac{s}{M_\tau^2}\right) \widehat{D}_{\rm pert}(a, L),
\end{equation}
 where the reduced function
 $\widehat{D}(s)\equiv D^{(1+0)}(s)-1$ is obtained
 by subtracting the dominant term  from the Adler function, {\em i.e.}  the logarithmic derivative of the polarization function, $D^{(1+0)}(s)\equiv - s\, {\rm d}\Pi^{(1+0)}(s)/{\rm d}s$ \cite{BrNaPi}. 

The function $\widehat{D}(s)$ depends only on the energy variable $s$, but for its pure perturbative part $\widehat{D}_{\rm pert}$ appearing  in (\ref{del0}) we emphasized also the formal dependence on the renormalization scale $\mu^2$,  entering through the  strong  coupling $\alpha_s(\mu^2)$ and the standard perturbative logarithms. Specifically, we define
\beq\label{aL}
 a\equiv \as(\mu^2)/\pi,\quad\quad  L\equiv\,\ln (-s/\mu^2).
\eeq
 In the so-called ``fixed-order perturbation theory", one chooses a fixed scale $\mu^2=M_\tau^2$ and the expansion of $\widehat{D}$ reads
\begin{equation}
\label{Ds}
 \widehat{D}_{\rm FOPT}(a, L) = \sum\limits_{n=1}^\infty a^n
\sum\limits_{k=1}^{n} k\, c_{n,k}\,L^{k-1} \,.
\end{equation}
 In the expansion above,  the leading coefficients $c_{n,1}$ are calculated from Feynman diagrams. The known coefficients $c_{n,1}$ are  (see \cite{BCK08} and references therein):
\beq\label{cn1}
c_{1,1}=1,\, c_{2,1}=1.640,\, c_{3,1}=6.371,\, c_{4,1}=49.076,
\eeq
and several estimates for the next coefficient $c_{5,1}$ were made recently \cite{BeJa, Beneke_Muenchen, Pich_Muenchen}.
The remaining coefficients $c_{n,k}$ for $k>1$  are determined from RG invariance and 
involve the coefficients $\beta_j$ appearing in the perturbation expansion of the RG  $\beta$-function 
\begin{equation}\label{beta}
\beta(a) \equiv \mu^2 \frac{\mathrm{d}a}{\mathrm{d}\mu^2} = -a^2 \sum_{k=0}^\infty \beta_k a^k.
\end{equation}
The $\beta$-function was calculated to four loops in the $\overline{{\rm MS}}$-renormalization scheme, the known coefficients being  (see \cite{LaRi,Czakon} for the calculation of $\beta_3$ and earlier references):
\beq\label{betaj}
 \beta_0=9/4,\, \beta_1=4,\, \beta_2= 10.0599,\,
\beta_3=47.228.
\eeq
 As remarked in \cite{dLP1}, due to the large imaginary part of the logarithm of $-s/ M_\tau^2$ along the circle $|s|= M_\tau^2$, the series (\ref{Ds})  is badly behaved especially near the timelike axis.  
This mandates one to search for expansions that would
be better behaved and would exhibit a smaller renormalization-scale dependence.  
The  ``contour-improved perturbation theory"  \cite{dLP1, Pivovarov:1991rh} is based on the RG-improved expansion, defined by the choice $\mu^2=-s$, when (\ref{Ds}) reduces to 
\beq\label{DsCI}
 \widehat{D}_{\rm CIPT}(\as(-s)/\pi, 0)= \sum\limits_{n=1}^\infty  c_{n,1} \left(\frac{\alpha_s(-s)}{\pi}\right)^n
\,.
\eeq
The main improvement comes from the treatment of the running coupling  $\alpha_s(-s)$, which  is determined by solving the RG Equation (\ref{beta}) numerically in an iterative way along the circle, starting with the input value $\alpha_s(M_\tau^2)$ at $s=-M_\tau^2 $. 

The expansions (\ref{Ds}) and (\ref{DsCI}) coincide formally as long as all the terms in the series are
retained (we ignore in this discussion the fact that the coefficients $c_{n,1}$ are known to increase as $n!$ and the series are actually divergent).  However, since the expansion coefficients are known
only up to a finite and not so large order, the series have to be
truncated at some order $n\leq N$. Then the expansions differ by terms of order $\alpha_s^{N+1}$, which may be substantial due to the relatively large value of the coupling at the low scale set by the mass of the $\tau$.  Therefore, the expansions lead to different values for $\delta^{(0)}$, this being the main source of error in the determination of  $\alpha_s(M_\tau^2)$ from the hadronic $\tau$-decays.

\section{Renormalization-Group Summation}\label{sec:RGR}
As suggested in \cite{Ahmady1, Ahmady2}, the FO expansion (\ref{Ds}) of the reduced Adler function can be written equivalently as 
\begin{equation}
\label{dseries}
 \widehat{D}_{\rm FOPT}(a, L) =
\sum_{n=1}^\infty a^n D_n (aL),
\end{equation}
 where the functions $D_n(u)$, depending on a single variable $u=aL$, are defined as 
\begin{equation}\label{Dn_def}
D_n (u) \equiv \sum_{k=n}^\infty (k-n+1)c_{k, k-n+1} u^{k-n}.
\end{equation}
As seen from the definition, the first function $D_1$ sums all the leading logarithms, the second function $D_2$ sums the next-to-leading logarithms, and so on. Thus, the suggestion was to effectively make a summation by collecting the aggregate coefficients of
the leading logarithms multiplied by fixed powers of the coupling
constant.  The attractive feature pointed out in \cite{Ahmady1, Ahmady2}, is that these functions can be obtained in a closed analytical form. We sketch below the derivation, which is based on RG invariance.

The Adler function defined by \eqn{dseries}, calculated in a fixed renormalization scheme,  is scale independent and  satisfies the RG equation 
\begin{equation}
\mu^2 \frac{\mathrm{d}}{\mathrm{d}\mu^2} \left\{  \widehat{D}_{\rm FOPT}(a, L) \right\} =
0,
\end{equation}
which can be written equivalently as
\begin{equation}
\label{pde2}
 \beta(a)
\frac{\partial \widehat{D}_{\rm FOPT}}{\partial a}-\frac{\partial \widehat{D}_{\rm FOPT}}{\partial L}  = 0.
\end{equation}
Using in this relation the expansion \eqn{Ds}  yields the following equation:
\bea
0= -\sum_{n=1}^\infty
\sum_{k=2}^n k (k-1) c_{n,k}  a^n L^{k-2}\nonumber\nonumber \\
- \left(\beta_0 a^2 + \beta_1 a^3 + \beta_2 a^4  + 
\ldots +\beta_l a^{l+2} +\ldots \right) \nonumber \nonumber \\ \times\sum_{n=1}^\infty
\sum_{k=1}^n n k c_{n,k}  a^{n-1} L^{k-1}\,.
\eea
By extracting the
aggregate coefficient of $a^n L^{n-p}$  one obtains the recursion formula
$(n \geq p)$
\begin{equation}\label{recursion}
0 = (n-p+2) c_{n, n-p+2}+ \sum_{\ell = 0} ^{p-2}  (n - \ell - 1) \beta_\ell c_{n - \ell - 1, n - p+1}.
\end{equation}
These relations are well known, and
in particular for $n\leq 4$ they coincide with the relations  given in Eq. (2.11) of \cite{BeJa}. 

Multiplying both sides of (\ref{recursion})  by $(n-p+1) u^{n-p}$   and summing from $n=p$ to
$\infty$, we obtain a set of first-order linear differential equation for the functions defined in (\ref{Dn_def}), written as
\begin{equation}
0  = \frac{\mathrm{d} D_{p-1}}{\mathrm{d}u} + u \sum_{\ell = 0}^{p-2} \beta_\ell \frac{\mathrm{d} D_{p-\ell - 1}}{\mathrm{d}u}
 +\sum_{\ell = 0}^{p-2} (p - \ell - 1) \beta_\ell D_{p - \ell - 1}.
\end{equation}
 Setting now $n = p - 1$  we write this set  as 
\beq\label{Dk_de}
\frac{\mathrm{d}D_n}{\mathrm{d}u} +  \sum_{\ell = 0}^{n-1} \beta_\ell \left( u \frac{\mathrm{d}}{\mathrm{d}u} + n - \ell \right) D_{n - \ell}=0,
\eeq
for $n\ge 1$, with the initial conditions $D_n (0) = c_{n,1}$ which follow from (\ref{Dn_def}). 

 The solution of the system (\ref{Dk_de}) can be found iteratively in an analytical closed form. It turns out that the solutions $D_n(u)$ depend on $u$ only through the variable $w=1 + \beta_0 u$. 
The expressions of $D_{n}(u)$ for $n=1,2,3,4$, written in terms of this variable and the coefficients $c_{n,1}$ and $\beta_k$,  are:
\bea\label{D12}
D_1 (u) && = \frac {c_{1, 1}} {w},\quad\quad\quad w=1 + \beta_0 u,\nonumber \\
D_2 (u) && =   \frac {c_{2, 1}} {w^2} - \frac {\beta_{1} c_{1,1} \ln w} {\beta_{0} w^2},
\eea
\begin{widetext}
\beq\label{D3}
D_3(u) =  \frac {(\beta_ {1}^2 - \beta_ {0}\beta_ {2}) c_ {1, 
    1} } {\beta_ {0}^2  w^{2} } +  \left [\frac {(-\beta_{1}^2 
 + \beta_{0} \beta_{2}) c_{1, 
     1}} {\beta_{0}^2} + 
 c_{3, 1} \right] w^{-3} 
 + \left [-\frac {\beta_{1} (\beta_{1} c_{1, 1} + 
      2\beta_{0} c_{2, 
        1}) \ln w} {\beta_{0}^2} 
+ \frac {\beta_{1}^2 c_{1, 1} \ln^2 w} {\beta_{0}^2}\right] w^{-3}. 
\eeq

\vspace{-0cm}
\bea
&&D_4 (u) =  -\frac {(\beta_{1}^3 - 
2\beta_{0} \beta_{1} \beta_{2} + \beta_{0}^2\beta_{3}) c_{1, 1}} {2\beta_ {0}^3} w^{-2}
-\left [\frac {\beta_{1} (-\beta_{1}^2 + \beta_{0} \beta_{2}) c_{1,1}} {\beta_{0}^3} + 
\frac {2 (-\beta_{1}^2 + \beta_{0} \beta_{2}) c_{2,1}} {\beta_{0}^2} \right] w^{-3} \nn
\nonumber \\
&& + \frac {2\beta_{1} (-\beta_{1}^2 + \beta_{0} \beta_{2}) c_{1, 1} 
\ln w} {\beta_{0}^3}w^{-3}
+  \left[\frac {(-\beta_{1}^3 + \beta_{0}^2\beta_{3}) c_{1,1}} 
{2\beta_{0}^3} + \frac {2 (-\beta_{1}^2 + \beta_{0} \
\beta_{2}) c_ {2,1}} {\beta_{0}^2} + c_{4,1}\right]w^{-4} 
\nonumber \\
&&-  \frac {\beta_{1} (-2 \beta_{1}^2 c_{1, 1} + 
3 \beta_{0} \beta_{2} c_{1, 1} + 
2 \beta_{0} \beta_{1} c_{2, 1} + 
3 \beta_{0}^2 c_{3, 1}) \ln w} {\beta_{0}^3}w^{-4}
+  \frac {\beta_{1}^2 (5\beta_{1} c_{1, 1} + 
6\beta_{0} c_{2,1}) \ln^2 w} {2\beta_{0}^3}w^{-4}
- \frac {\beta_{1}^3 c_{1,1} \ln^3 w} {\beta_{0}^3} w^{-4}.\nn
\eea
\end{widetext}
In \cite{Ahmady1,Ahmady2} similar differential equations were solved for $n\leq 4$ for several observables, including the cross section of $e^+ e^-$ annihilation into hadrons, whose expansion in QCD is related to the expansion of the Adler function in which we are interested. The functions $D_n(u)$ given above coincide actually with those calculated in \cite{Ahmady2}. For the applications made in this work and possible further studies, we have derived the expressions of $D_n$ up to  $n =10$.  The solutions depend on the coefficients  $c_{n,1}$ and the coefficients $\beta_k$  of the expansion (\ref{beta}) of the $\beta$-function.  For consistency, to each Feynman diagram order $n$ we use the expansion of the $\beta$-function to the same order. The complete expressions are rather lengthy. They simplify considerably if we insert the known numerical values of the coefficients $c_{n,1}$ for $n\le 4$ given in (\ref{cn1}), and of the coefficients $\beta_k$ for $k\leq 3$ given in (\ref{betaj}). The corresponding expressions, which depend on the arbitrary coefficients $c_{n,1}$ for $5\leq n\leq 10$, and  $\beta_k$ for $4\le k\le 9$,  are listed in Appendix.

We shall use in what follows the truncated FOPT improved by renormalization-group summation (RGS) written  as
\begin{equation}\label{Ds1}
D_{\rm IFOPT}(a, L) = \sum_{n=1}^{N} a^n D_n (aL).
\end{equation}

\section{Discussion}\label{sec:disc}
\begin{figure}[thb]
 	\begin{center}\vspace{0.5cm}
 	 \includegraphics[width = 7.3cm]{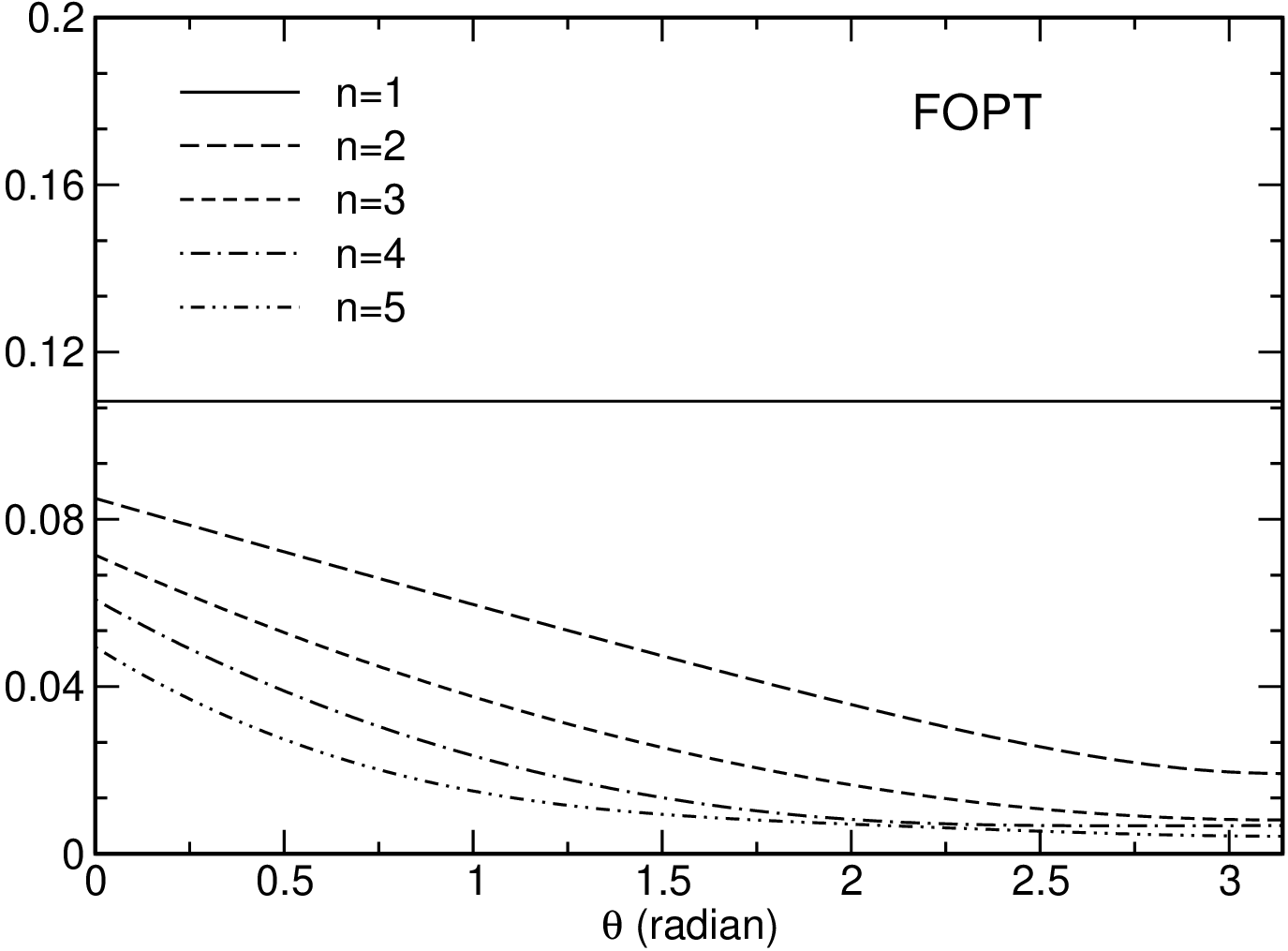}
	\caption{Modulus of the perturbative terms of the standard FO expansion (\ref{Ds})  along the circle $s=M_\tau^2 \exp(i\theta)$.}
	\label{fig_fopt}
 	\end{center}\vspace{0.0cm}
\end{figure}

\begin{figure}[thb]
 	\begin{center}\vspace{0.0cm}
 	 \includegraphics[width = 7.3cm]{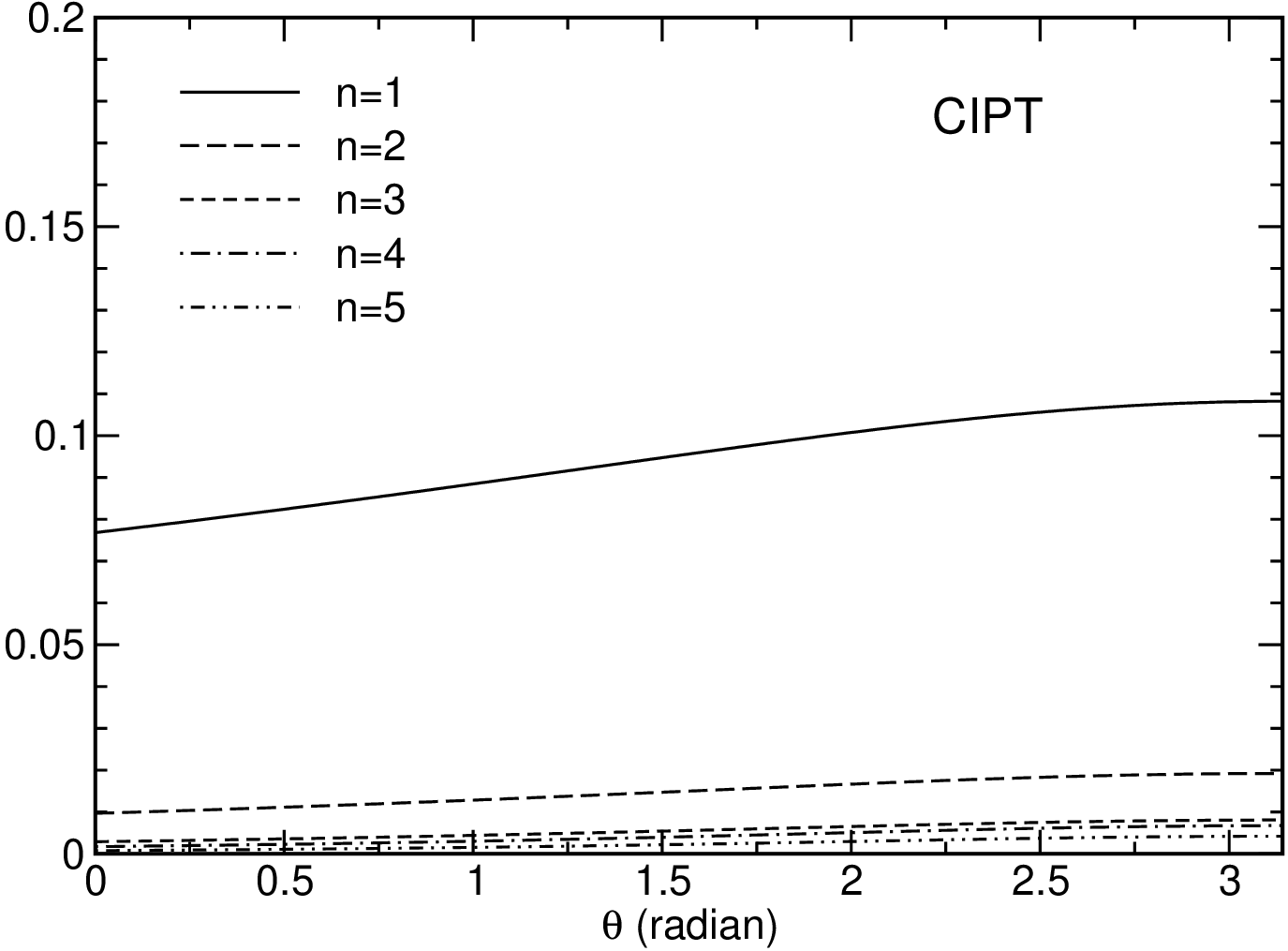}
	\caption{Modulus of the perturbative terms of the CI expansion (\ref{DsCI})  along the circle $s=M_\tau^2 \exp(i\theta)$. }
	\label{fig_cipt}
 	\end{center}\vspace{0.0cm}
\end{figure}

\begin{figure}[thb]
 	\begin{center}\vspace{0.0cm}
 	 \includegraphics[width = 7.3cm]{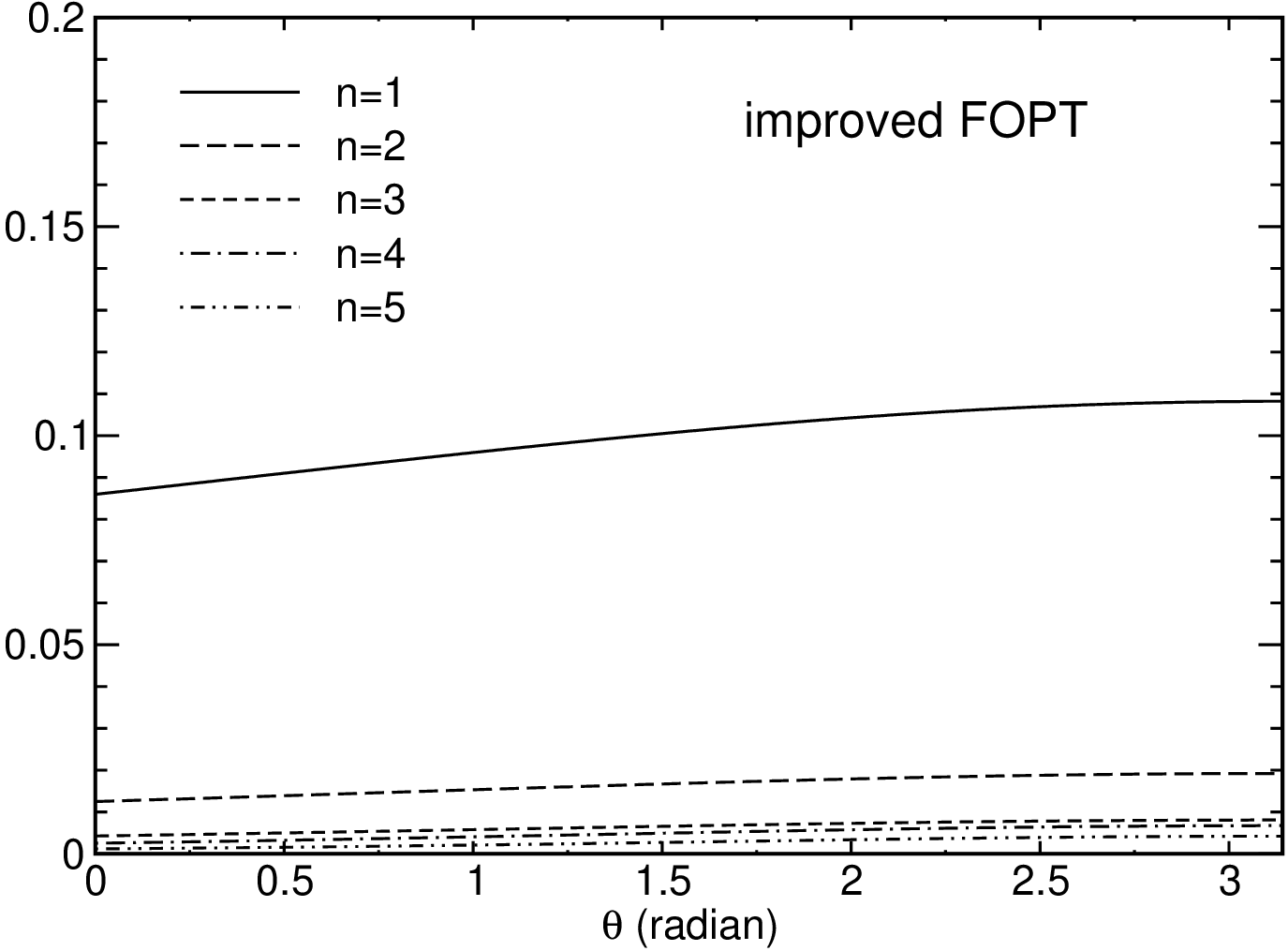}
	\caption{Modulus of the perturbative terms of the improved FO expansion (\ref{Ds1})  along the circle $s=M_\tau^2 \exp(i\theta)$. }
	\label{fig_ifopt}
 	\end{center}\vspace{0.0cm}
\end{figure}

In this section we shall investigate the properties of the new expansion (\ref{Ds1}) in the complex $s$-plane, along the circle $s=M_\tau^2 \exp(i\theta)$.   For comparison, we plot in  Figs. \ref{fig_fopt}-\ref{fig_ifopt} the modulus of each successive term of order $n\leq 5$ of the standard FOPT expansion (\ref{Ds}), the  CIPT expansion  (\ref{DsCI}) and the RGS improved FOPT expansion (\ref{Ds1}), respectively. For convenience, we have taken $\alpha_s(M_\tau^2)=0.34$. For $n=5$  we
have used the expression of $D_5$ given in the Appendix, with the estimate $c_{5,1}=283$ from \cite{BeJa,Beneke_Muenchen} and setting $\beta_4=0$. 


\begin{figure}[thb]
 	\begin{center}\vspace{0.0cm}
 	 \includegraphics[width = 7.3cm]{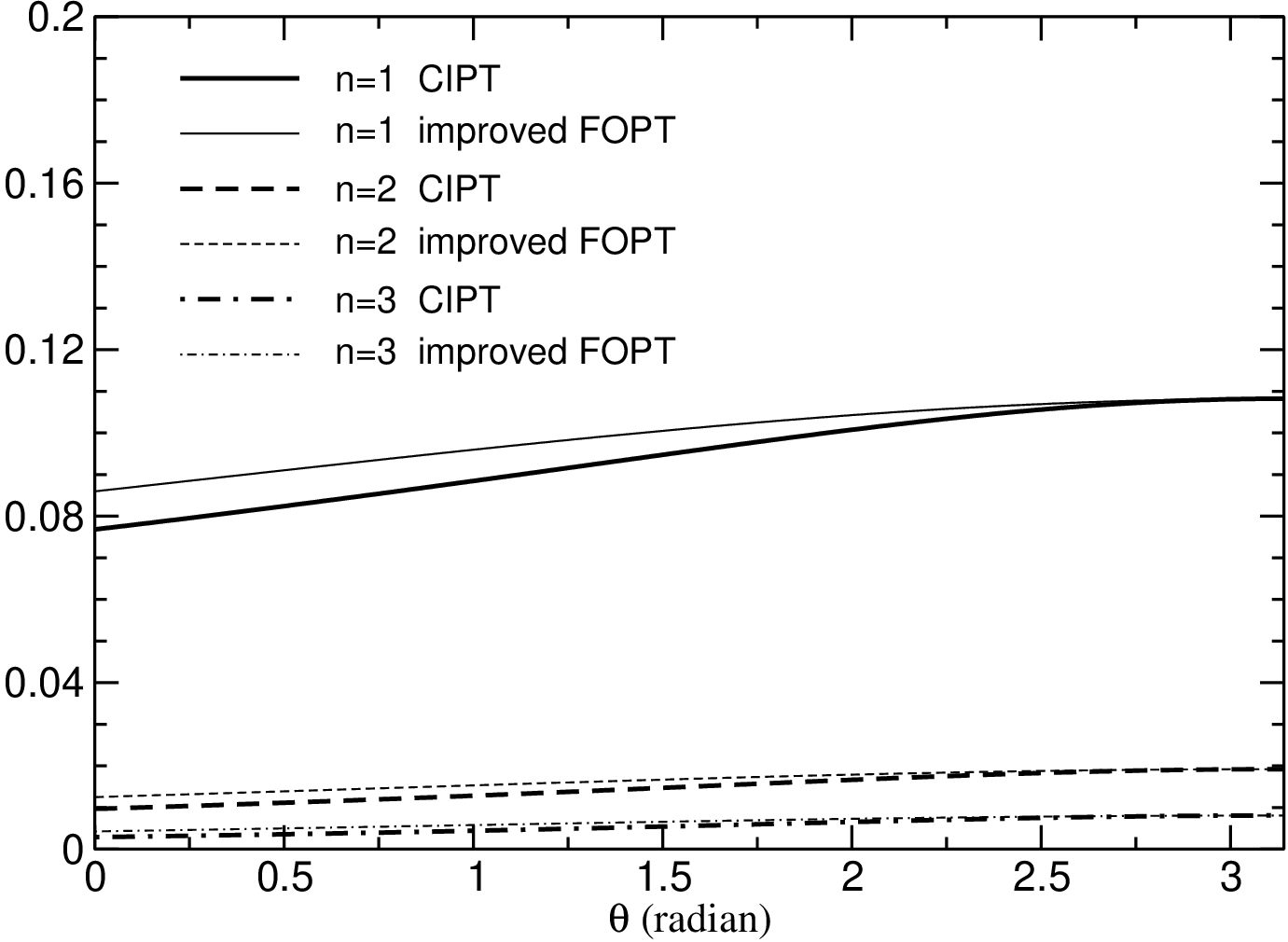}
	\caption{Comparison of the CI expansion (\ref{DsCI}) and the improved FO expansion (\ref{Ds1}) along the circle $s=M_\tau^2 \exp(i\theta)$. }
	\label{fig_ci_if}
 	\end{center}\vspace{0.0cm}
\end{figure}
\begin{figure}[thb]
 	\begin{center}\vspace{0.0cm}
 	 \includegraphics[width = 7.3cm]{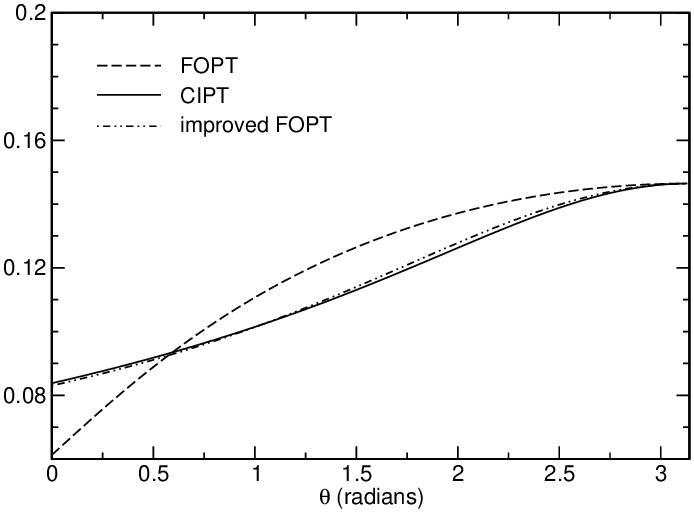}
	\caption{ Adler function expansions (\ref{Ds}),  (\ref{DsCI}) and  (\ref{Ds1}), summed up to the order $N=5$, along the circle $s=M_\tau^2 \exp(i\theta)$. }
	\label{fig_comp}
 	\end{center}\vspace{0.0cm}
\end{figure}
From Fig. \ref{fig_fopt} it is seen that the higher-order terms are large  near the timelike axis ($\theta=0$). This shows the slow convergence of the standard FOPT in this region, where the logarithm defined in (\ref{aL}) acquires a large imaginary part.  As discussed in  \cite{dLP1}, the reason is the poor convergence, especially near the timelike axis, of the expansion of $\alpha_s(-s)$ in powers of $\alpha_s(M_\tau^2)$, which is used in passing from the renormalization-group improved expansion (\ref{DsCI}) to the fixed-order expansion (\ref{Ds1}).
In contrast,  Fig. \ref{fig_cipt} shows that in CIPT the higher terms are much smaller, {\em i.e.} the expansion has a good convergence along the whole circle.   
As seen from Fig. \ref{fig_ifopt},   the RGS improved FOPT expansion (\ref{Ds1}) has a behavior similar to that of CIPT: the series is stable along the circle and the higher order terms are very  small.  Thus, although it depends explicitly only on the coupling at a fixed scale, the expansion (\ref{Ds1}) shares the good qualities of the CI expansion along the circle, as seen from  Fig. \ref{fig_ci_if}, where we simultaneously plot the first three terms for the two expansions.

In order to see the difference between CIPT and the FOPT improved by RGS, it is useful to look at the leading term, with $n=1$.  In the CI expansion (\ref{DsCI}) this term is $c_{1,1} \alpha_s(-s)/\pi$, where the coupling is calculated as the numerical solution of the  RG Eq. (\ref{beta}), keeping four terms in the expansion of the $\beta$-function. On the other hand, using  (\ref{Ds1}) and (\ref{D12}) we write the leading term of the RGS improved expansion as $c_{1,1} a/(1+\beta_0  a\ln(-s/M_\tau^2))$ where $a=\alpha_s(M_\tau^2)/\pi$. This is actually the exact solution of the RG Eq. (\ref{beta}) to one loop, written in terms of the input $\alpha_s(M_\tau^2)$. The similar behavior of the curves corresponding to $n=1$  in Fig. \ref{fig_ci_if} shows that the effect of the higher order terms in the expansion of the $\beta$-function is small. Moreover, the smallness of the next terms of the expansion (\ref{Ds1}) proves that the summation of the leading logarithms is very efficient also to higher orders.

Figure \ref{fig_comp} shows the behavior along the circle of the Adler function given by the first $N=5$ terms in the expansions (\ref{Ds}), (\ref{DsCI}) and (\ref{Ds1}), respectively. The new FO expansion improved by RGS is very similar to the CI expansion, as expected from the previous figures.

\hspace{0.0cm}
\begin{table}
\begin{tabular}{c c c c}
\hline \vspace{0.1cm}
$ $ & $\delta^{(0)}_{\rm FOPT}$ & $\delta^{(0)}_{\rm CIPT}$   & $\delta^{(0)}_{\rm IFOPT}$ \nonumber \\ \hline 
$ N=1 $ &~~ 0.1082~~ & ~~0.1479~~ & ~~ 0.1455 ~~ \nonumber \\ 
$ N=2 $ & 0.1691  & 0.1776 &  0.1797 \nonumber \\ 
$ N=3 $ & 0.2025 & 0.1898 &  0.1931 \nonumber \\ 
$ N=4 $ & 0.2199  & 0.1984   & 0.2024\nonumber \\ 
$ N=5 $ & 0.2287   & 0.2022     &  0.2056 \nonumber \\  \hline 
\end{tabular}
\caption{Predictions of $\delta^{(0)}$ by the standard FOPT, CIPT and  the RGS improved FOPT,
 for various truncation orders $N$.}
\label{tab:table1}
\end{table}

By inserting the FOPT, CIPT, and RGS improved FOPT expansions (\ref{Ds}), (\ref{DsCI}) (\ref{Ds1}), respectively,  truncated at some $N$, into the definition (\ref{del0}) of $\delta^{(0)}$, we obtain the corresponding values denoted as 
$\delta^{(0)}_{\rm FOPT}$, $\delta^{(0)}_{\rm CIPT}$  and $\delta^{(0)}_{\rm IFOPT}$ respectively. In  Table \ref{tab:table1} we list these values for various truncation orders $N\leq 5$, using in the calculation the standard value $\as(M_{\tau}^2)=0.34$.
 As remarked already,  CIPT shows a faster convergence compared to the standard FOPT.  To order $N=4$, the difference between FOPT and CIPT is $0.0215$, which, as remarked, is  the dominant theoretical  uncertainty in the extraction of $\alpha_s$ from the hadronic  
$\tau$ decay rate.
On the other hand, for $N=4$, the difference between the results of the RGS improved FOPT and the standard  FOPT is $0.01754$, and the difference from  the RGS improved FOPT and CIPT is $0.0039$, which confirms that the new expansion gives results close to those of the CIPT. 
For $N=5$, using the estimate $c_{5,1}=283$ from \cite{BeJa}, we find that the RGS improved FOPT differs from FOPT by $0.0232$, and from  CIPT by $0.0035$.

It is of interest to see whether this behavior is preserved to higher orders. To this end we consider a class of physical models of the Adler function used for testing various expansions in \cite{BeJa,CaFi2009,Jamin_Manchester,CaFi_Manchester,CaFi2011}.
 
\begin{figure}[thb]
 	\begin{center}\vspace{0.5cm}
 	 \includegraphics[width = 7.3cm]{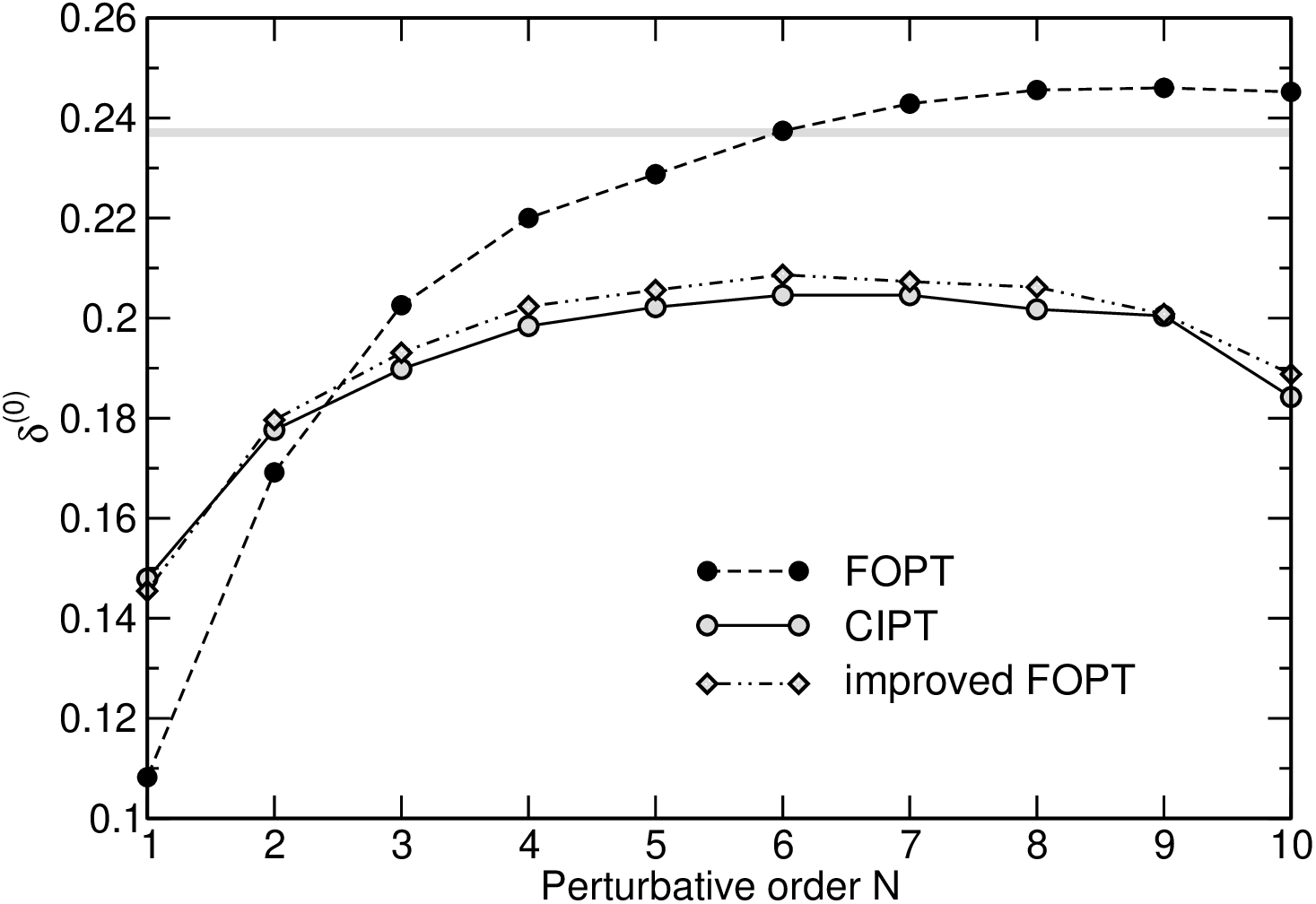}
	\caption{Dependence of $\delta^{(0)}$ in FOPT, CIPT and RGS 
improved FOPT on the truncation order $N$ in the Beneke and Jamin model \cite{BeJa}. The gray band is the exact value obtained with the expressions (\ref{eq:pv})-(\ref{eq:dBJ}).}
	\label{fig_del}
 	\end{center}\vspace{0.0cm}
\end{figure}

In particular, we consider the model proposed in \cite{BeJa}, where the Adler function is defined in terms of its Borel transform $B(u)$  by the principal value prescription
\beq\label{eq:pv}
\wh D(s)=\frac{1}{\beta_0}\,{\rm PV} \,\int\limits_0^\infty  e^{-\frac{u}{\beta_0 a(-s)}} \, B(u)\, {\rm d} u,
\eeq
where the function $B(u)$ is expressed in terms of a few ultraviolet (UV) and infrared (IR) renormalons 
\beq\label{eq:BBJ}
B_{\rm BJ}(u)=B_1^{\rm UV}(u) +  B_2^{\rm IR}(u) + B_3^{\rm IR}(u) +d_0^{\rm PO} + d_1^{\rm PO} u.
\eeq
In  \cite{BeJa} these terms were written  as 
\bea\label{eq:BIRUV}
B_p^{\rm IR}(u)= 
\frac{d_p^{\rm IR}}{(p-u)^{\gamma_p}}\,
\Big[\, 1 + \tilde b_1 (p-u) + \ldots \,\Big], \nonumber \\
B_p^{\rm UV}(u)=\frac{d_p^{\rm UV}}{(p+u)^{\bar\gamma_p}}\,
\Big[\, 1 + \bar b_1 (p+u)  +\ldots \,\Big],
\eea
where most of the parameters were obtained by imposing RG invariance at four loops. Finally, the free parameters of the model, {\em i.e.} the residues $d_1^{\rm UV}, d_2^{\rm IR}$ and  $d_3^{\rm IR}$ of the first renormalons and the coeficients $d_0^{\rm PO}, d_1^{\rm PO}$ of the polynomial in (\ref{eq:BBJ}),  were fixed  by the requirement of reproducing the perturbative coefficients $c_{n,1}$ for $n\le 4$ from (\ref{cn1}) and the estimate $c_{5,1}=283$, and read:
\begin{equation}\label{eq:dBJ}
d_1^{\rm UV}=-\,1.56\times 10^{-2},~~~
d_2^{\rm IR}=3.16, ~~~
d_3^{\rm IR}=-13.5,\nn\nonumber \\[-1mm]
\end{equation}
\begin{equation}\label{eq:dBJ1}
d_0^{\rm PO}=0.781, ~~~
d_1^{\rm PO}=7.66\times 10^{-3}. 
\end{equation}
Then all the higher order coefficients $c_{n,1}$ are fixed and exhibit a factorial increase, showing that the perturbative series of the Adler function is  divergent. We list below the values, given in \cite{BeJa},  which we used in our analysis 
\bea\label{cn11}
c_{5,1}&\!\!=\!\!&283,\, c_{6,1}=3275,\,  c_{7,1}=1.88\times 10^4, \nonumber \\
 c_{8,1}&\!\!=\!\!& 3.88\times 10^5, \, c_{9,1}= 9.19\times 10^5,\,c_{10,1}= 8.37\times 10^7.\nn
\eea
In Fig. \ref{fig_del}, we show the exact value of $\delta^{(0)}$ obtained with the above model, and the dependence of the truncation order $N$ for the three expansions considered: standard FOPT, standard CIPT and RGS improved FOPT. 
As in the previous figures we have used as input $\alpha_s(M_\tau^2)=0.34$. For the RGS improved FOPT we have used the expressions of $D_n$ given in the appendix, setting  $\beta_k=0$ for $k\ge 4$ as in the previous similar analyses  of higher order expansions \cite{BeJa, CaFi2009,Jamin_Manchester,CaFi2011}.

The figure shows that the FOPT improved by RGS gives results close to the CIPT predictions at all orders up to $N=10$. 
In fact, as remarked in \cite{BeJa}, for this particular model the standard FO expansion describes better than the CIPT the ``true" function. Indeed, as seen in Fig. \ref{fig_del}, up to $N=10$ the predictions of the CI expansion stay below the true result, and in fact never approach it (for higher truncation orders $N$  all the three expansions start to show big oscillations, due to the divergent character of the series). 

 We mention however that, as discussed in \cite{Jamin_Manchester,CaFi2011}, for other models the CI expansion may give better results than the standard FOPT at low orders. In particular, this is true for models with a  residue  $d_2^{\rm IR}$ of the first IR renormalon smaller than the value quoted in (\ref{eq:dBJ}). In our work we investigated numerically several such models, the conclusion being that in all cases the   fixed-order expansion improved by RG-summation gives results close to those obtained by the contour-improved expansion.

\section{Determination of $\alpha_s(M_\tau^2)$}\label{sec:alphas}
In this section we shall use the RGS improved FO expansion (\ref{Ds1}) for a  determination of $\alpha_s(M_\tau^2)$ in the $\overline{\rm MS}$ scheme.
 We use as input the  phenomenological value of the pure perturbative correction to the hadronic $\tau$ width estimated recently in \cite{Beneke_Muenchen} from the ALEPH data 
\begin{equation}
\label{delph}
 \delta^{(0)}_{\rm phen}=0.2037\pm 0.0040_{exp}\pm 0.0037_{\rm PC},
\end{equation}
where the first error is experimental and the second reflects the uncertainty of the higher order terms (``power corrections") in the OPE. 
 We note that a similar value, $\delta^{(0)}_{\rm phen}=0.2038\pm 0.0040$, is quoted also in the recent review \cite{Pich_Muenchen}. On the other hand, the recent fits of the moments of the OPAL spectral function in the frame of OPE for the Adler function including duality violating terms \cite{DV} suggest that the error of the nonperturbative contribution may be larger. As the issue is still under investigation, we stick in our analysis  to the input (\ref{delph}), used in several recent determinations \cite{CaFi2009, Beneke_Muenchen, CaFi2011}.

 For the theoretical evaluation of  $\delta^{(0)}$ from (\ref{del0}) we apply the improved FO expansion (\ref{Ds1}) truncated at $N=5$,  choosing the scale as $\mu^2 = \xi M_{\tau}^2$ with $\xi= 1\pm 0.5$. We have used the functions $D_n$ for $n\le 5$ given in the Appendix, taking as input in $D_5$ the conservative estimates $ c_{5,1}= 283 \pm 283$ \cite{BeJa,Beneke_Muenchen,CaFi2011} and  $\beta_4=0 \pm \beta_3^2/\beta_2$, as in \cite{Davier2008, Pich_Muenchen,CaFi2011}.  With this input we obtained from the phenomenological value (\ref{delph}) the result
\bea\label{alpha}
\alpha_s(M_\tau^2)&=&0.3378  \pm  0.0046_{\rm exp}  \pm 0.0042_{\rm PC} ~^{+0.0062}_{- 0.0072}(c_{5,1})\nn
\nonumber \\&&^{+ 0.0005}_{-0.0004 }{(\rm scale)} \pm ^{+ 0.000085}_{-0.000082}{(\rm \beta_4)}.
\eea
In this result the first two errors are due to the corresponding uncertainties of $\delta^{(0)}_{\rm phen}$ given in (\ref{delph}), the third one reflects the uncertainty of the coefficient  $c_{5,1}$ with the very conservative range adopted above, the fourth is due to scale variation, and the last one shows the effect of the truncation of the $\beta$-function expansion. One may note the very small sensitivity of $\alpha_s(M_\tau^2)$ on the variation of the scale, and a relatively large contribution of the uncertainty of the five loop coefficient  $c_{5,1}$, a feature noticed also in the standard CIPT analyses \cite{Pich_Muenchen, Pich_Manchester} and in the CI expansions improved by the conformal mappings of the Borel plane \cite{CaFi2009,CaFi2011}.

Combining in quadrature the errors given in (\ref{alpha}), we write (\ref{alpha}) as
\begin{equation}\label{aver1}
\alpha_s(M_\tau^2)= 0.338 \pm 0.010.
\end{equation}
We mention that for the same input (\ref{delph}) the standard FOPT and CIPT give, respectively, 
\bea\label{FOCI}
\alpha_{s}(M_{\tau}^2)&=& 0.320^{+0.012}_{-0.007}, \quad\quad\quad{\rm FOPT}, \nonumber \\ \alpha_{s}(M_{\tau}^2)&=& 0.342 \pm  0.012, \quad\quad{\rm CIPT}.
\eea
For comparison we mention also the value  $\alpha_{s}(M_{\tau}^2)= 0.320^{+0.019}_{-0.014}$, obtained recently in \cite{CaFi2011} with the same input (\ref{delph})  and an improved CI expansion based on the analytic continuation in the Borel plane.

\section{Summary and Conclusion}\label{sec:conc}

In this paper we have applied the method of explicit summation of all RG-accessible logarithms proposed in \cite{Ahmady1,Ahmady2}  to the perturbative expansion of the Adler function relevant for the extraction of $\alpha_s$ from $\tau$ hadronic decays.   We thus refer to the resulting scheme as ``FOPT improved  by RG-summation", or  ``improved FOPT".   The work is motivated by the well-known  discrepancy between the predictions of $\alpha_s(M_\tau^2)$ from the standard fixed-order and RG-improved expansions.  As this discrepancy has to do with the behavior of the perturbative expansion of the Adler function  along the contour  involved in the integral (\ref{del0}), especially near the timelike axis, it was of interest to see whether a more general fixed-order expansion  can be found, with good convergence properties along the contour. While the method proposed in \cite{Ahmady1,Ahmady2} was applied to several other observables, its properties in the complex energy plane were not investigated until now.

As mentioned earlier, several modifications of the standard FO and CI perturbative expansion were recently proposed  and applied to the Adler function, for the determination of the strong coupling from $\tau$ decays \cite{CaFi2009,CaFi2011,CLMV1}. The present approach exploits RG invariance in a complete way, summing in analytical closed expressions all the terms that can be calculated to a definite Feynman diagram order.  Of course, the truncated expansions of the different summations  differ among each other by terms of order $\alpha_s^{N+1}$, which may be quite important  at the relatively low scale relevant in $\tau$ decays. Moreover, the actual  differences depend on the detailed form of including known information on the higher order terms.  Therefore, our study contributes to the assessment of the ambiguities of the perturbation expansion of the Adler function in the complex plane and the theoretical error of $\alpha_s(M_\tau^2)$.

 The main result of the paper is that the summation of leading logarithms provides a systematic expansion with good convergence properties in the complex plane, including the critical region near the timelike region.  
 By summing up pieces of the standard fixed-order series (\ref{Ds1}) into the functions $D_n$ defined in (\ref{Dn_def}), the new expansion (\ref{Ds1}) is no longer plagued by  large imaginary parts of the logarithms, responsible for the poor convergence of FOPT  along the contour.

On the other hand,  the results of the new expansion are close to those obtained with the CI expansion (\ref{DsCI}), which was to be expected since both implement RG invariance.   As discussed in Sec. \ref{sec:disc}, the behavior of the new expansion along the circle $|s|=M_\tau^2$ is similar to that of CIPT.
 However, the two expansions are not identical:  CIPT uses the exact solution of the RG equation to four loops, found numerically by an iterative integration along the circle,  while the new expansion involves only  expressions written in an analytically closed form valid along the whole integration contour, thereby avoiding numerical integration. 

Using as input the recent estimates \cite{Beneke_Muenchen,Pich_Muenchen} of the perturbative correction to the $\tau$ hadronic width,  the new expansion (\ref{Ds1}) to five loops leads to the value (\ref{aver1}) for $\alpha_s(M_\tau^2)$ in the $\overline{\rm MS}$ scheme. The result is situated  between the predictions of FOPT and CIPT given in (\ref{FOCI}), closer to the latter. 
 We emphasize that the error given in (\ref{aver1}) reflects in particular the uncertainty of the nonperturbative contribution to the hadronic width of $\tau$ quoted in  (\ref{delph}).  Of course, a definite answer to the issue of these corrections requires the simultaneous extraction of $\alpha_s$ and the power corrections from the moment analysis of the spectral function, accounting also for the duality violating terms, as in the recent work \cite{DV}. The improved FO expansion investigated here, having the advantage that is written in an analytically closed form to each order, could be useful in such an analysis in the future. 

\vspace{0.cm}
\subsection*{Acknowledgements} I.C. acknowledges support from CNCS in the program Idei, Contract No. 121/25.10.2011.

\vskip0.cm
\appendix*
\section{Expressions of the functions $D_n$}
We give the expressions of the functions $D_{n}(u)$, $n=1,2, \ldots ,10$ in a readily readable form
using the known numerical values of the coefficients  $c_{n,1}$ for $1 \leq n \leq 4$ from (\ref{cn1}), and of $\beta_j$ for  $0\leq j \leq 3 $ from (\ref{betaj}). The higher coefficients  $c_{n,1}$ for $n\ge 5$ are arbitrary. For generality, at each order $n$ we include the higher loop coefficients $\beta_j$ for $j\ge 4$ up to the corresponding order.

As remarked in Sec. \ref{sec:RGR}, the functions $D_{n}(u)$ depend only on the variable 
\beq
 w =  1 + 9/4 u.\eeq
The explicit expressions are:



\begin{gather}
\begin{split}
D_{1} (u) &= w^{-1}.
\end{split}
\end{gather}


\begin{gather}
 \begin{split}
D_2 (u) = & ( 1.64 - 1.778 \ln w )w^{-2}.
 \end{split}
\end{gather}

\begin{widetext}
\vskip-0.4cm
\begin{gather}
\begin{split}
D_3(u) = & -1.311 w^{-2} 
+ \left (7.682 - 8.992 \ln w + 3.160\ln^2 w \right) w^{-3}. 
\end{split}
\end{gather}


\begin{gather}
 \begin{split}
D_4 (u) = & -5.356 w^{-2}
+ \left(-6.629 + 4.660 \ln w \right) w^{-3} 
\\
+ & \left(61.061 -56.954 \ln w
+ 29.596 \ln^2 w - 5.619 \ln^3 w \right) w^{-4}.
\end{split}
\end{gather}

\vspace{0.01cm}

\begin{gather}
 \begin{split}
D_5 (u) = & \left(20.740 - 0.148  \beta_{4} \right) w^{-2}
+ \left(-25.371 + 19.043\ln w \right) w^{-3}
\\
+ & \left(-41.986 + 43.637\ln w - 12.426 \ln^2 w \right) w^{-4}
\\
+ & \left(46.618 + 0.148 \beta_{4} + c_{5,1} - 535.458\ln w \right) w^{-5}
\\
+ & \left(255.117\ln^2 w - 80.143 \ln^3 w + 9.989 \ln^4 w \right) w^{-5}
 \end{split}
\end{gather}

\vspace{-0cm}

\begin{gather}
 \begin{split}
D_6 (u) = & \left(-8.802 + 0.395 \beta_{4} - 0.111 \beta_{5}  \right) w^{-2}
\\
+ & \left(118.935 - 0.749 \beta_{4} + \left(-73.7407 
+ 0.527 \beta_{4}\right)\ln w \right) w^{-3}
\\
+ & \left(-155.498 + 169.168\ln w - 50.782\ln^2 w \right) w^{-4}
\\
+ & \left(-394.738 + 376.142\ln w - 
 177.243\ln^2 w + 29.455 \ln^3 w \right) w^{-5}
\\
+ & \left(440.104 + 0.354 \beta_{4} + 0.111 \beta_{5} 
+  c_{6,1} + (-1366.3 - 1.317 \beta_{4} - 8.889 c_{5,1})\ln w \right) w^{-6}
\\
+ & \left( 2833.36\ln^2 w - 898.378\ln^3 w + 
 195.853\ln^4 w - 17.758 \ln^5 w \right) w^{-6}
 \end{split}
\end{gather}
\begin{align}
D_7 (u) & = \left(1.850 - 0.048 \beta_{4} + 0.316 \beta_{5} 
- 0.089  \beta_{6} \right) w^{-2}
\nonumber \\
& + \left(-70.196 + 2.386 \beta_{4} - 
 0.562 \beta_{5} + (31.297 - 1.405 \beta_{4} 
+ 0.395 \beta_{5})\ln w \right) w^{-3}
\nonumber \\
& + \left(793.632 - 
 4.746 \beta_{4} + (-765.413 + 4.933 \beta_{4})\ln w + (196.642 - 
     1.405 \beta_{4})\ln^2 w \right) w^{-4}
\nonumber \\
& + \left(-1474.52 + 1406.51\ln w - 691.764\ln^2 w + 120.371\ln^3 w \right) w^{-5}
\nonumber \\
& + \left(-1007.23 - 0.971 \beta_{4} - 6.553 c_{5,1} \right) w^{-6}
\nonumber \\
& + \left(4177.48\ln w - 
1986.84\ln^2 w + 577.528\ln^3 w - 65.455 \ln^4 w \right) w^{-6}
\nonumber \\
& + \left(1756.47 + 3.378 \beta_{4} + 0.246 \beta_{5} 
+ 0.089 \beta_{6} + 6.553 c_{5,1} + c_{7,1}\right) w^{-7}
\nonumber \\
& + \left((-7123.42 - 6.119 \beta_{4} - 1.185 \beta_{5} 
- 15.803 c_{5,1} - 10.667 c_{6,1})\ln w \right) w^{-7}
\nonumber \\
& + \left((12324 + 7.023 \beta_{4} + 47.407 c_{5,1})\ln^2 w 
- 11671.3\ln^3 w \right) w^{-7}
\nonumber \\
& + \left(2743.86\ln^4 w - 449.389\ln^5 w 
+ 31.569 \ln^6 w  \right) w^{-7}
\end{align}
\begin{align}
D_8 (u) & = \left(-194.169 + 1.242 \beta_{4} - 0.040 \beta_{5} 
+ 0.263 \beta_{6} - 0.074  \beta_{7} \right) w^{-2}
\nonumber \\
& + \left(-189.727+ 0.309 \beta_{4} +1.890 \beta_{5} -0.449 \beta_{6} \right) w^{-3}
\nonumber \\
& + \left(-550.03 + 16.293 \beta_{4} - 
 3.559 \beta_{5} \right) w^{-4}
\nonumber \\
& + \left((430.016 - 15.225 \beta_{4} 
+ 3.699 \beta_{5})\ln w + (-83.458 + 
     3.746 \beta_{4} - 1.053 \beta_{5})\ln^2 w \right) w^{-4}
\nonumber \\
& + \left(7347.75 - 44.622 \beta_{4} + (-7004.34 + 42.519 \beta_{4})\ln w \right) w^{-5}
\nonumber \\
& + \left((3071.05 - 20.036 \beta_{4})\ln^2 w 
+ (-466.114 + 3.329 \beta_{4})\ln^3 w \right) w^{-5}
\nonumber \\
& + \left(-2576.43 - 3.967 \beta_{4} - 26.779 c_{5,1} + 15607.3\ln w \right) w^{-6}
\nonumber \\
& + \left(-7480.95\ln^2 w + 2263.66\ln^3 w - 267.492\ln^4 w \right) w^{-6}
\nonumber \\
& + \left(-5251.36 - 4.511 \beta_{4} - 0.874 \beta_{5} - 11.650 c_{5,1} 
- 7.863 c_{6,1}\right) w^{-7}
\nonumber \\
& + \left((18170.5 + 10.355 \beta_{4} + 69.897 c_{5,1})\ln w - 25812.\ln^2 w + 
 8091.04\ln^3 w \right) w^{-7}
\nonumber \\
& + \left(-1656.44\ln^4 w + 139.637 \ln^5 w  \right) w^{-7}
\nonumber \\
& + \left(1413.96 + 35.257 \beta_{4} + 2.584 \beta_{5} 
+ 0.186 \beta_{6} \right) w^{-8}
\nonumber \\
+& \left( 0.074 \beta_{7} + 
 38.429 c_{5,1} + 7.863 c_{6,1} + c_{8,1} \right) w^{-8}
\nonumber \\
& + \left((-34522.1 - 52.922 \beta_{4} - 5.167 \beta_{5} - 1.106 \beta_{6} 
- 109.64 c_{5,1} - 
    18.963 c_{6,1} - 12.444 c_{7,1})\ln w \right) w^{-8}
\nonumber \\
& + \left((66232.9 + 50.564 \beta_{4} + 7.374 \beta_{5} + 182.606 c_{5,1} 
+ 66.370 c_{6,1})\ln^2 w \right) w^{-8}
\nonumber \\
& + \left((-71870.8 - 29.134 \beta_{4} - 196.653 c_{5,1})\ln^3 
+ 41188.6\ln^4 w \right) w^{-8}
\nonumber \\
& + \left(-7628.07\ln^5 w + 988.189\ln^6 w 
- 56.123 \ln^7 w \right) w^{-8}
\end{align}
\vspace{-0.0cm}
\begin{align}
  D_9 (u) & = \left(395.544 - 10.428 \beta_{4} + 0.028 \beta_{4}^2 
+ 1.064 \beta_{5} - 0.034 \beta_{6} + 
 0.226 \beta_{7} -0.063 \beta_{8}  \right) w^{-2}
\nonumber \\
& + \left(-462.494 - 3.971 \beta_{4} + 0.022 \beta_{4}^2 
+ 0.160 \beta_{5} + 1.565 \beta_{6} - 
 0.375 \beta_{7} \right) w^{-3}
 \nonumber \\
& + \left((690.377 - 4.415 \beta_{4} + 0.142 \beta_{5} 
- 0.936 \beta_{6} + 0.263 \beta_{7})\ln w \right) w^{-3}
\nonumber \\
& + \left(-1810.96 + 2.864 \beta_{4} + 12.852 \beta_{5} - 2.848 \beta_{6} - 
 2.220 \times  10^{-16} \beta_{8}\right) w^{-4}
\nonumber \\
& + \left((1000.18 - 1.345 \beta_{4} - 12.076 \beta_{5} + 2.960 \beta_{6})\ln w \right) w^{-4}
\nonumber \\
& + \left((17.544 - 0.455 \beta_{4} + 2.997 \beta_{5} 
- 0.843 \beta_{6})\ln^2 w\right) w^{-4}
\nonumber \\
& + \left(-5114.46 + 150.336 \beta_{4} - 33.466 \beta_{5} + 1.421 \times  10^{-14} \beta_{6} - 
 3.553 \times  10^{-15} \beta_{7} - 8.882 \times  10^{-16} \beta_{8}\right) w^{-5}
\nonumber \\
& + \left((4675.79 - 142.926 \beta_{4} + 31.890 \beta_{5})\ln w 
+ (-1677.31 + 60.792 \beta_{4} - 
     15.027 \beta_{5})\ln^2 w\right) w^{-5}
 \nonumber \\
& + \left((197.827 - 8.879 \beta_{4} + 2.497 \beta_{5})\ln^3 w\right) w^{-5}
\nonumber \\
& + \left(28164.8 - 98.496 \beta_{4} - 0.110 \beta_{4}^2 + 3.411\times  10^{-13} \beta_{5} - 
 2.842\times  10^{-14} \beta_{6}\right) w^{-6}
\nonumber \\
& + \left(1.421 \times  10^{-14} \beta_{7} + 3.553 \times  10^{-15} \beta_{8} + 
 103.698 c_{5,1} - 0.741 \beta_{4} c_{5,1} \right) w^{-6}
\nonumber \\
& + \left((-77765.5 + 472.226 \beta_{4}) \ln w 
+ (36590. - 224.594 \beta_{4}) \ln^2 w \right) w^{-6}
\nonumber \\
& + \left((-9928.07 + 65.284 \beta_{4}) \ln^3 w 
+ (1035.81 - 7.399 \beta_{4}) \ln^4 w \right) w^{-6}
\nonumber \\
& + \left(-14763 - 14.620 \beta_{4} - 2.665 \times  10^{-15} \beta_{4}^2 
- 3.571 \beta_{5} \right) w^{-7}
 \nonumber \\
& + \left(1.421 \times  10^{-13 } \beta_{6} + 1.776 \times  10^{-15} \beta_{8} - 
 21.844 c_{5,1} - 32.135 c_{6,1} \right) w^{-7}
\nonumber \\
& + \left((55228.2 + 42.318 \beta_{4} + 285.647 c_{5,1}) \ln w - 96538.4 \ln^2 w + 
 30623.2 \ln^3 w\right) w^{-7}
\nonumber \\
& + \left(-6511.98 \ln^4 w + 570.649 \ln^5 w\right) w^{-7}
\nonumber \\
& + \left(-25449.6 - 39.014 \beta_{4} - 8.882 \times  10^{-16} \beta_{4}^2 
- 3.809 \beta_{5} - 
 0.815 \beta_{6} + 7.105 \times  10^{-15} \beta_{7} \right) w^{-8}
\nonumber \\
& + \left(1.776 \times  10^{-15} \beta_{8} - 80.826 c_{5,1} 
- 13.979 c_{6,1} - 9.174 c_{7,1}\right) w^{-8}
\nonumber \\
& + \left((97653.3 + 74.551 \beta_{4} + 10.873 \beta_{5} + 269.233 c_{5,1} 
+ 97.856 c_{6,1}) \ln w\right) w^{-8}
\nonumber \\
& + \left((-158949. - 64.432 \beta_{4} - 434.915 c_{5,1}) \ln^2 w 
+ 121456 \ln^3 w \right) w^{-8}
\nonumber \\
& + \left(-28116.9\ln^4 w + 
 4370.93\ln^5 w - 289.617 \ln^6 w  \right) w^{-8}
 \nonumber \\
& + \left(19040.2 + 13.328 \beta_{4} + 0.060 \beta_{4}^2 
+ 26.770 \beta_{5} + 2.132 \beta_{6} + 
 0.149 \beta_{7} \right) w^{-9}
 \displaybreak[0] \nonumber \\
& + \left(0.063 \beta_{8} - 1.028 c_{5,1} 
+ 0.741 \beta_{4} c_{5,1} + 46.115 c_{6,1} + 
 9.174 c_{7,1} + c_{9,1}\right) w^{-9}
 \nonumber \\
& + \left((-81482.4 - 595.514 \beta_{4} 
- 45.930 \beta_{5} - 4.615 \beta_{6} - 1.054 \beta_{7} \right) w^{-9}
\nonumber \\
& + \left (- 741.46 c_{5,1} - 145.547 c_{6,1} - 22.124 c_{7,1} 
- 14.222 c_{8,1}) \ln w \right) w^{-9}
\nonumber \\
& + \left((363238. + 466.224 \beta_{4} + 49.8562 \beta_{5} 
+ 7.86612 \beta_{6} + 1104.3 c_{5,1} + 
    252.84 c_{6,1} + 88.4938 c_{7,1}) \ln^2 w \right) w^{-9}
\nonumber \\
& + \left((-441763. - 291.503 \beta_{4} - 34.961 \beta_{5} - 1215.29 c_{5,1} - 
    314.645 c_{6,1}) \ln^3 w \right) w^{-9}
\nonumber \\
& + \left((328765. + 103.587 \beta_{4} + 699.21 c_{5,1}) \ln^4 w 
- 130720. \ln^5 w \right) w^{-9}
\nonumber \\
& + \left(19838.1\ln^6 w - 2107.52\ln^7 w 
+ 99.775 \ln^8 w \right) w^{-9}
\end{align}
\begin{align}
 D_{10} (u) & =  \left(11.539 + 14.642 \beta_{4} - 0.132 \beta_{4}^2 
- 9.125 \beta_{5} + 0.049 \beta_{4} \beta_{5} \right) w^{-2} 
\nonumber \\
& +  \left(0.931 \beta_{6} - 0.030 \beta_{7} + 0.198 \beta_{8} 
- 0.056 \beta_{9} \right) w^{-2}
\nonumber \\
& +  \left(2124.59 - 36.489 \beta_{4} 
+ 0.026 \beta_{4}^2 - 2.506 \beta_{5} + 0.033 \beta_{4} \beta_{5} \right) w^{-3}
\nonumber \\
& +  \left( 0.089 \beta_{6} + 1.336 \beta_{7} - 0.321 \beta_{8} \right) w^{-3}
\nonumber \\
& +  \left((-1406.38 + 37.077 \beta_{4} - 0.1 \beta_{4}^2 - 3.784 \beta_{5} 
+ 0.122 \beta_{6} - 
    0.803 \beta_{7} + 0.226 \beta_{8}) \ln w \right) w^{-3}
\nonumber \\
& +  \left(-680.52 - 60.419 \beta_{4} + 0.206 \beta_{4}^2 + 1.535 \beta_{5} - 
 2.220 \times  10^{-16} \beta_{4} \beta_{5} \right) w^{-4}
\nonumber \\
& +  \left(10.619 \beta_{6} - 2.373 \beta_{7} 
+ 2.220 \times  10^{-16} \beta_{9} \right) w^{-4}
\nonumber \\
& +  \left((3693.97 + 13.328 \beta_{4} - 0.117 \beta_{4}^2 
- 0.599 \beta_{5} - 10.012 \beta_{6} + 
    2.466 \beta_{7}) \ln w \right) w^{-4}
\nonumber \\
& +  \left((-1841.01 + 11.774 \beta_{4} - 0.379 \beta_{5} + 2.497 \beta_{6} - 
    0.702 \beta_{7}) \ln^2 w \right) w^{-4}
\nonumber \\
 &+  \left(-15549.1 + 19.182 \beta_{4} - 3.553 \times  10^{-15} \beta_{4}^2 
+ 118.701 \beta_{5} \right) w^{-5}
\nonumber \\
& +  \left(-1.776\times 10^{-15 } \beta_{4} \beta_{5} - 26.773 \beta_{6} - 
 2.842\times  10^{-14} \beta_{7} - 1.776\times  10^{-15} \beta_{9} \right) w^{-5}
\nonumber \\
& +  \left((14656. - 22.760 \beta_{4} - 112.864 \beta_{5} 
+ 25.512 \beta_{6}) \ln w \right) w^{-5}
\nonumber \\
& +  \left((-3525. + 3.973 \beta_{4} + 48.266 \beta_{5} 
- 12.021 \beta_{6}) \ln^2 w \right) w^{-5}
\nonumber \\
& +  \left((-41.587 + 1.077 \beta_{4} - 7.103 \beta_{5} 
+ 1.998 \beta_{6}) \ln^3 w \right) w^{-5}
 \nonumber \\
& +  \left( -34138. + 645.226 \beta_{4} + 0.293 \beta_{4}^2 
- 85.394 \beta_{5} - 0.082 \beta_{4} \beta_{5} - 
 4.547 \times  10^{-13} \beta_{6}\right) w^{-6}
\nonumber \\
& + \left(5.684 \times  10^{-14} \beta_{7} - 3.553  \times  10^{-15} \beta_{9} - 
 44.011 c_{5,1} + 1.975 \beta_{4} c_{5,1} - 0.556 \beta_{5} c_{5,1} \right) w^{-6}
\nonumber \\
& + \left((53774.4 - 1590.41 \beta_{4} + 354.169 \beta_{5}) \ln w 
+ (-23763.2 + 743.3 \beta_{4} - 
     168.446 \beta_{5}) \ln^2 w \right) w^{-6}
\nonumber \\
& + \left((5321.51 - 195.91 \beta_{4} + 48.963 \beta_{5}) \ln^3 w 
+ (-439.615 + 19.731 \beta_{4} - 
     5.549 \beta_{5}) \ln^4 w \right) w^{-6}
\nonumber \\
& + \left(133168. - 491.03 \beta_{4} - 0.510 \beta_{4}^2 + 13.826 \beta_{5}
 - 0.099 \beta_{4} \beta_{5} + 
 9.095  \times  10^{-13} \beta_{6} \right) w^{-7}
\nonumber \\
& + \left(5.684  \times  10^{-14} \beta_{7} + 394.93 c_{5,1} - 1.317 \beta_{4} c_{5,1} + 
 124.437 c_{6,1} - 0.889 \beta_{4} c_{6,1} \right) w^{-7}
 \nonumber \\
& + \left((-438674. + 1890.14 \beta_{4} + 1.171 \beta_{4}^2 - 1106.11 c_{5,1} + 
    7.901 \beta_{4} c_{5,1}) \ln w \right) w^{-7}
\nonumber \\
& + \left((479798. - 2917.82 \beta_{4}) \ln^2 w + (-147748. 
+ 914.619 \beta_{4}) \ln^3 w \right) w^{-7}
\nonumber \\
& + \left((28316.3 - 187.245 \beta_{4}) \ln^4 w 
+ (-2209.73 + 15.785 \beta_{4}) \ln^5 w \right) w^{-7}
\nonumber \\
& + \left(-68009.4 - 131.957 \beta_{4} + 2.842 \times 10^{-14} \beta_{4}^2 - 11.56 \beta_{5} - 
 3.553  \times  10^{-15} \beta_{4} \beta_{5} - 3.333 \beta_{6} \right) w^{-8}
\nonumber \\
& + \left(-2.842  \times 10^{-13} \beta_{7} - 1.421 \times  10^{-14} \beta_{8} - 
 7.105  \times  10^{-15} \beta_{9} \right) w^{-8}
\nonumber \\
& + \left(- 231.071 c_{5,1} - 21.060 c_{6,1} - 
 37.491 c_{7,1} \right) w^{-8}
\nonumber \\
& + \left((281901. + 257.166 \beta_{4} + 44.434 \beta_{5} + 779.652 c_{5,1} 
+ 399.906 c_{6,1}) \ln w \right) w^{-8}
\nonumber \\
& + \left((-515266. - 263.312 \beta_{4} - 1777.36 c_{5,1}) \ln^2 w 
+ 454897. \ln^3 w \right) w^{-8}
\nonumber \\
& + \left(-106849. \ln^4 w + 17222.1 \ln^5 w - 1183.57 \ln^6 w \right) w^{-8}
\nonumber \\
& + \left(-60068.5 - 439.01 \beta_{4} - 3.553  \times  10^{-14} \beta_{4}^2 
- 33.859 \beta_{5} - 
 3.553  \times  10^{-15} \beta_{4} \beta_{5} \right) w^{-9}
\nonumber \\
& + \left(- 3.402 \beta_{6} 
- 0.777 \beta_{7} 1.421  \times  10^{-14} \beta_{8} 
+ 1.776  \times  10^{-15} \beta_{9} - 
 546.601 c_{5,1} \right) w^{-9}
 \nonumber \\
& + \left(-107.297 c_{6,1} - 16.309 c_{7,1} - 10.485 c_{8,1} 
- 8.882  \times  10^{-16} \beta_{5} c_{5,1} \right) w^{-9}
\nonumber \\
& + \left(-107.297 c_{6,1} - 16.309 c_{7,1} - 10.485 c_{8,1} \right) w^{-9}
\displaybreak[0] \nonumber \\
& + \left((535556. + 687.396 \beta_{4} + 73.508 \beta_{5} 
+ 11.598 \beta_{6} )\ln w \right) w^{-9}
\nonumber \\
& + \left( 1628.16 c_{5,1} + 
    372.785 c_{6,1} + 130.475 c_{7,1}) \ln w  \right) w^{-9}
\nonumber \\
& + \left((-976999. - 644.685 \beta_{4} - 77.318 \beta_{5} 
-2687.73 c_{5,1} - 695.865 c_{6,1}) \ln^2 w \right) w^{-9}
\nonumber \\
& + \left((969457. + 305.455 \beta_{4} + 2061.82 c_{5,1}) \ln^3 w 
-481830. \ln^4 w + 87747.5 \ln^5 w 
-10875.6 \ln^6 w + 588.427 \ln^7 w  \right) w^{-9}
\nonumber \\
& + \left(43141.6 + 479.855 \beta_{4} + 0.118 \beta_{4}^2 
+ 8.381 \beta_{5} + 0.099 \beta_{4} \beta_{5} + 
 21.868 \beta_{6} + 1.844 \beta_{7} \right) w^{-10}
\nonumber \\
& + \left(0.124 \beta_{8} + 0.056 \beta_{9} + c_{10,1} + 426.754 c_{5,1} 
- 0.658 \beta_{4} c_{5,1} \right) w^{-10}
\nonumber \\
& + \left(0.556 \beta_{5} c_{5,1} + 3.919 c_{6,1} 
+ 0.889 \beta_{4} c_{6,1} + 53.80 c_{7,1} + 
 10.485 c_{8,1} \right) w^{-10}
\nonumber \\
& + \left((-449500. - 1271.94 \beta_{4} - 0.953 \beta_{4}^2 
- 509.968 \beta_{5} - 42.321 \beta_{6} - 
  4.255 \beta_{7} - 1.016 \beta_{8}) \ln w \right) w^{-10}
\displaybreak[0] \nonumber \\
& + \left((-1301.71 c_{5,1} - 11.852 \beta_{4} c_{5,1} 
- 996.586 c_{6,1} - 186.115 c_{7,1} - 
    25.284 c_{8,1} - 16. c_{9,1}) \ln w \right) w^{-10}
\nonumber \\
& + \left((1.298  \times  10^6 + 5592.95 \beta_{4} + 456.071 \beta_{5} 
+ 50.903 \beta_{6}) \ln^2 w \right) w^{-10}
\nonumber \\
& + \left((8.428 \beta_{7} + 7894.87 c_{5,1} + 1613.87 c_{6,1} + 334.31 c_{7,1} + 
    113.778 c_{8,1}) \ln^2 w \right) w^{-10}
\nonumber \\
& + \left((-2.723  \times 10^6 - 3004.75 \beta_{4} 
- 328.052 \beta_{5} - 41.953) \ln^3 w \right) w^{-10}
\nonumber \\
& + \left((\beta_{6} - 8050.1 c_{5,1} - 1907.85 c_{6,1} 
- 471.967 c_{7,1}) \ln^3 w \right) w^{-10}
 \nonumber \\
& + \left((2.352  \times  10^6 + 1350.17 \beta_{4} + 139.842 \beta_{5} 
+ 6104.22 c_{5,1} + 
   1258.58 c_{6,1}) \ln^4 w \right) w^{-10}
\nonumber \\
& + \left((-1.284 \times  10^6 - 331.478 \beta_{4} - 2237.47 c_{5,1}) \ln^5 w + 
 383853. \ln^6 w \right) w^{-10}
\nonumber \\
& + \left(-49091.\ln^7 w + 
 4392.42\ln^8 w -177.377 \ln^9 w \right) w^{-10}.
\end{align}
\end{widetext}

\end{document}